\documentclass[useAMS,usenatbib,twocolumn]{aastex61_myfix}
\usepackage{natbib,graphics,epsfig}
\usepackage{amsmath}
\usepackage{amssymb}
\usepackage{textcomp}
\usepackage{graphicx}
\usepackage{color}
\usepackage{hyperref}
\usepackage{txfonts}

\usepackage{graphicx}
\usepackage{color}
\usepackage[T1]{fontenc}
\usepackage{aecompl}

\submitjournal{ApJL}

\shorttitle{Wandering SMBHs}
\shortauthors{Tremmel et al.}

\begin{document}

\title{Wandering Supermassive Black Holes in Milky Way Mass Halos}

\correspondingauthor{Michael Tremmel}
\email{michael.tremmel@yale.edu}

\author[0000-0002-4353-0306]{Michael Tremmel}
\affil{Yale Center for Astronomy \& Astrophysics, Physics Department, P.O. Box 208120, New Haven, CT 06520, USA}

\author{Fabio Governato}
\affil{Astronomy Department, University of Washington, Box 351580, Seattle, WA, 98195-1580}

\author{Marta Volonteri}
\affil{Sorbonne Universit\`{e}s, UPMC Univ Paris 6 et CNRS, UMR 7095, Institut d`Astrophysique de Paris, 98 bis bd Arago, 75014 Paris, France}

\author{Andrew Pontzen}
\affil{Department of Physics and Astronomy, University College London, 132 Hampstead Road, London, NW1 2PS, 
United Kingdom}

\author{Thomas R. Quinn}
\affil{Astronomy Department, University of Washington, Box 351580, Seattle, WA, 98195-1580}




\begin{abstract}
We present a self-consistent prediction from a large-scale cosmological simulation for the population of `wandering' supermassive black holes (SMBHs) of mass greater than $10^6$ M$_{\odot}$ on long-lived, kpc-scale  orbits within Milky Way (MW)-mass galaxies. We extract a sample of MW-mass halos from the {\sc Romulus25} cosmological simulation \citep{tremmel17}, which is uniquely able to capture the orbital evolution of SMBHs during and following galaxy mergers. We predict that such halos, regardless of recent merger history or morphology, host an average of  $5.1 \pm 3.3$ SMBHs, including their central black hole, within 10 kpc from the galactic center and an average of $12.2 \pm 8.4$ SMBHs total within their virial radius, not counting those in satellite halos. Wandering SMBHs exist within their host galaxies for several Gyrs, often accreted by their host halo in the early Universe. We find, with $>4\sigma$ significance, that wandering SMBHs are preferentially found outside of galactic disks.
\end{abstract}

\keywords{Galaxy: kinematics and dynamics --- quasars: supermassive black holes}

\section{Introduction}



New evidence for the existence of massive black holes ($> 10^4$ M$_{\odot}$) near the Galactic center \citep[][but see also \citet{ravi17}]{oka17,tsuboi17} raises fundamental questions about the formation of supermassive black holes (SMBHs) and their evolution within galaxies. Combined with the growing number of observed offset and dual active galactic nuclei \citep[e.g.][]{comerford11,comerford14,barrows16,barrows17}, this establishes the idea that non-central SMBHs may be relatively common and potentially observable not only in the Milky Way (MW) but also in other galaxies. Previous studies have suggested that the orbital decay experienced by SMBHs accreted onto a galaxy through minor mergers can be stalled, creating a population of SMBHs that fail to reach the galactic center within a Hubble time \citep{G94,schneider02,volonteriOffCenBH05,bellovaryBH10,tremmel15,tremmel18,dosopoulou17,dvorkin17}. \citet{tremmel18} find that only a fraction of mergers involving MW-mass halos result in the formation of close SMBH pairs. Rather, many galaxy mergers (and the majority of minor mergers) result in SMBHs that are deposited on wide orbits following the disruption of their host galaxy during the interaction. Because these  wandering SMBHs often come from smaller galaxies, their masses are likely close to their initial mass. Due to their connection to a previous population of satellite galaxies, future observations of wandering SMBHs may provide important insight into how often, at what mass, and in what halos SMBHs are formed.

The {\sc Romulus} cosmological simulations \citep{tremmel17} are uniquely capable of tracking the orbital evolution of SMBHs within their host galaxies to sub-kpc accuracy \citep{tremmel15,tremmel18}. SMBHs are also seeded in the early Universe based on gas properties. This allows SMBHs to exist in smaller halos and at earlier times compared with more common approaches. In this Letter we use the {\sc Romulus25} simulation to self-consistently predict the average number of wandering SMBHs and their dynamics within MW-mass galaxies.

In \S2 we discuss the simulations, the sub-grid physics implemented for SMBHs, and our halo selection criteria. In \S3 we discuss our results predicting the population of wandering SMBHs in MW-mass galaxies, which are summarized in \S4.

\section{The Simulations}

{\sc Romulus25} is a a $25^3$ Mpc$^3$ uniform volume simulation run with a $\Lambda$CDM cosmology following the most recent results from Planck \citep[$\Omega_0=0.3086$, $\Lambda=0.6914$, h$=0.67$, $\sigma_8=0.77$;][]{planck16}, a Plummer equivalent force softening of $250$ pc (a $350$ pc spline force softening is used), and mass resolution for dark matter and gas of $3.39 \times 10^5$ and $2.12 \times 10^5$ M$_{\odot}$ respectively. The simulation was run using the new Tree + SPH code, {\sc ChaNGa} \citep{changa15}, including models for  a cosmic UV background, star formation, `blastwave' supernovae (SN) feedback, low temperature metal cooling  \citep{wadsley04,wadsley08,Stinson06,shen10}, as well as a novel implementation of SMBH formation, growth and dynamics \citep{tremmel15,tremmel17}. {\sc Romulus25} reproduces  $z=0$ empirical relations between halo mass, galaxy stellar mass, and SMBH mass. It also results in realistic cosmic star formation and SMBH growth histories \citep{tremmel17}.

SMBHs of mass $10^6$ M$_{\odot}$ are seeded from gas in rapidly collapsing, pristine regions capable of quickly producing a very massive black hole. More than 85\% of SMBHs are seeded within the first Gyr of the simulation without \textit{a priori} assumptions regarding their halo occupation. This results in SMBHs being seeded in $10^8-10^{10}$ M$_{\odot}$ halos, with their occupation evolving such that only $\sim10\%$ of $10^{10}$ M$_{\odot}$ halos host SMBHs at $z = 0$. This seeding method allows for a more complete census of SMBHs throughout each galaxy's merger history. 

Once seeded, SMBHs are allowed to grow by accreting gas via a modified Bondi-Hoyle prescription that utilizes the local \textit{resolved} kinematics of gas to account for angular momentum support. Free parameters associated with sub-grid models for SMBH and stellar physics were constrained by an extensive parameter optimization. For more details we refer the reader to \citet{tremmel17}.

Crucial to this work, the orbital evolution of SMBHs in {\sc Romulus25} is tracked down to sub-kpc scales by utilizing the sub-grid model presented in \citet{tremmel15} to account for unresolved dynamical friction. This method has been explicitly shown to result in realistic SMBH orbital evolution \citep{tremmel15,tremmel18}. SMBHs form close pairs when they are within two softening lengths ($0.7$ kpc) of one another and are considered relatively bound ($\frac{1}{2}\Delta \textbf{v} < \Delta \textbf{a} \cdot \Delta \textbf{r}$, where $\Delta \textbf{v}$, $\Delta \textbf{a}$, and $\Delta \textbf{r}$ are the relative velocity, acceleration, and distance vectors between two SMBHs). When this occurs, the individual kinematics are no longer followed and the two SMBHs act as a single object with the sum of the two masses and the same total momentum. We therefore consider SMBHs in the central $0.7$ kpc of the halo to be central SMBHs and all others to be `wanderers'. While this work has been motivated by recent claims of massive, wandering black holes in the Milky Way, our simulations are not tuned to reproduce these results, nor do we model SMBHs smaller than $10^6$ M$_{\odot}$. Rather, the orbital evolution of SMBHs in {\sc Romulus25} is purely a prediction of the simulation. 


We use the Amiga Halo Finder \citep{knollmann09} to extract individual halos from the volume. SMBH positions are taken relative to the center of the halo and their velocities relative to the center of mass velocity of the halo's inner 1 kpc. We define `Milky Way-like' to include halos with total mass between $5\times10^{11}$ and $2\times10^{12}$ M$_{\odot}$. {\sc Romulus25} contains 26 such halos at $z=0$, excluding all satellites. Halos in this mass range are large enough to have rich merger histories and substantial populations of wandering SMBHs. Smaller galaxies have quieter merger histories and fewer wandering SMBHs while there are much fewer larger galaxies in {\sc Romulus25}. The similarity to the MW and Andromeda is also convenient because, due to their proximity, they may be promising initial laboratories for wandering SMBH studies.

The stellar masses of these halos are similar to that of the Milky Way ($\sim 10^{10.15}-10^{10.85}$ M$_{\odot}$) after applying a factor to account for observational limitations \citep{munshi13}. We focus on SMBHs within the inner 10 kpc of halos, as this represents a region dominated by the central galaxy (for reference the MW disk is $\sim10$ kpc in radius).


\section{The Population of Wandering SMBHs in Milky Way-mass Halos}

\begin{figure}
\centering
\includegraphics[trim=15mm 5mm 10mm 20mm, clip, width=90mm]{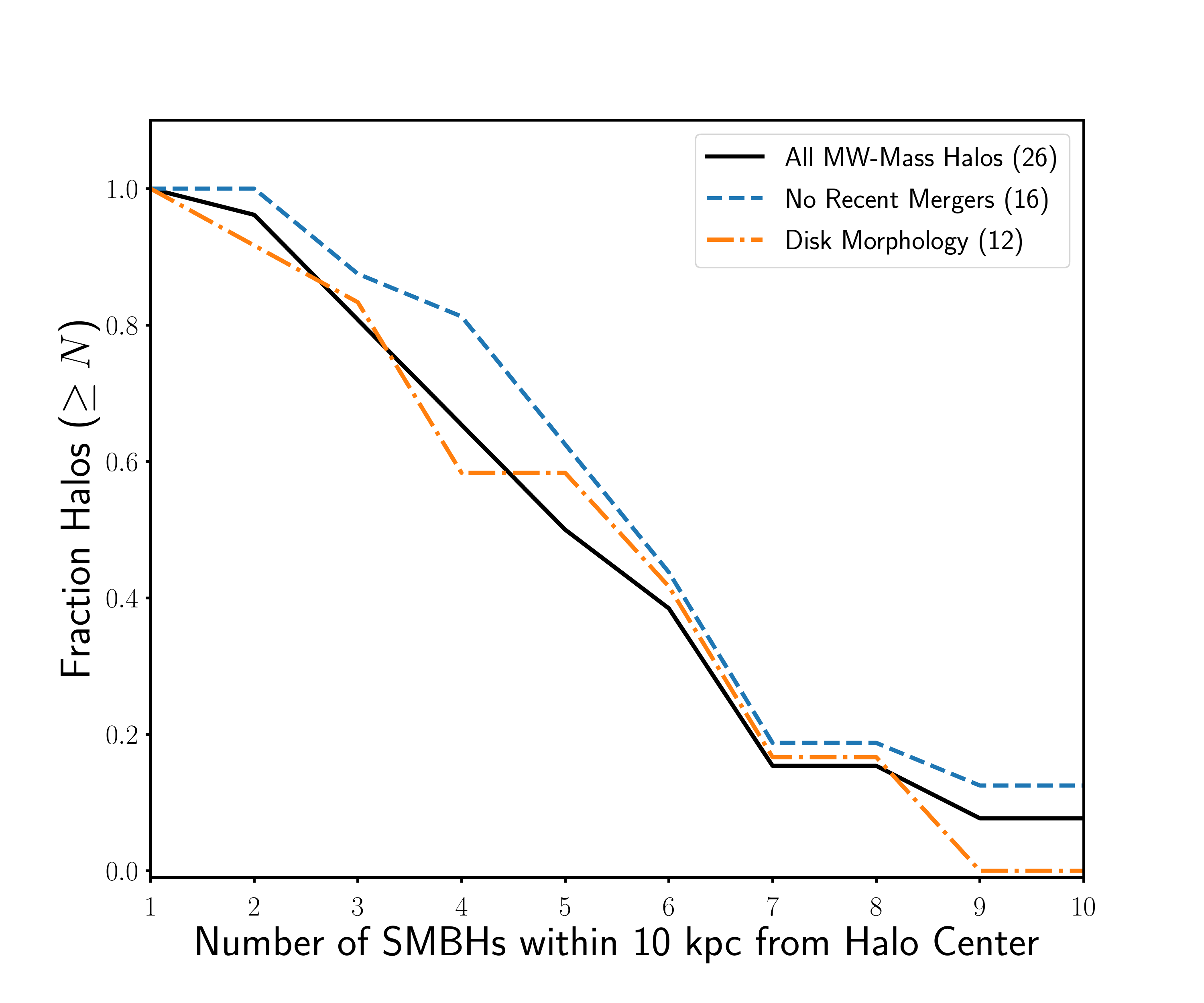}
\caption{{\sc Wandering SMBHs near the center of MW-mass Halos}. The cumulative fraction of MW-mass halos in {\sc Romulus25} as a function of the number of SMBHs they host, including central SMBHs. All halos host at least one SMBH within 10 kpc from halo center but the majority host more than that. The black line represents the entire sample, the blue dashed line the sub-sample without major mergers sinze $z=0.75$, and the orange dot-dash line the sub-sample visually categorized as having a disk morphology. Hosting several wandering SMBHs is the norm for MW-mass halos (only 1 of the 26 in {\sc Romulus25} does not have any). The number of wandering SMBHs is insensitive to morphology or recent merger history.}
\label{number_bhs} 
\end{figure}

 \begin{figure*}
\centering
\includegraphics[trim=15mm 17mm 20mm 15mm, clip, width=150mm]{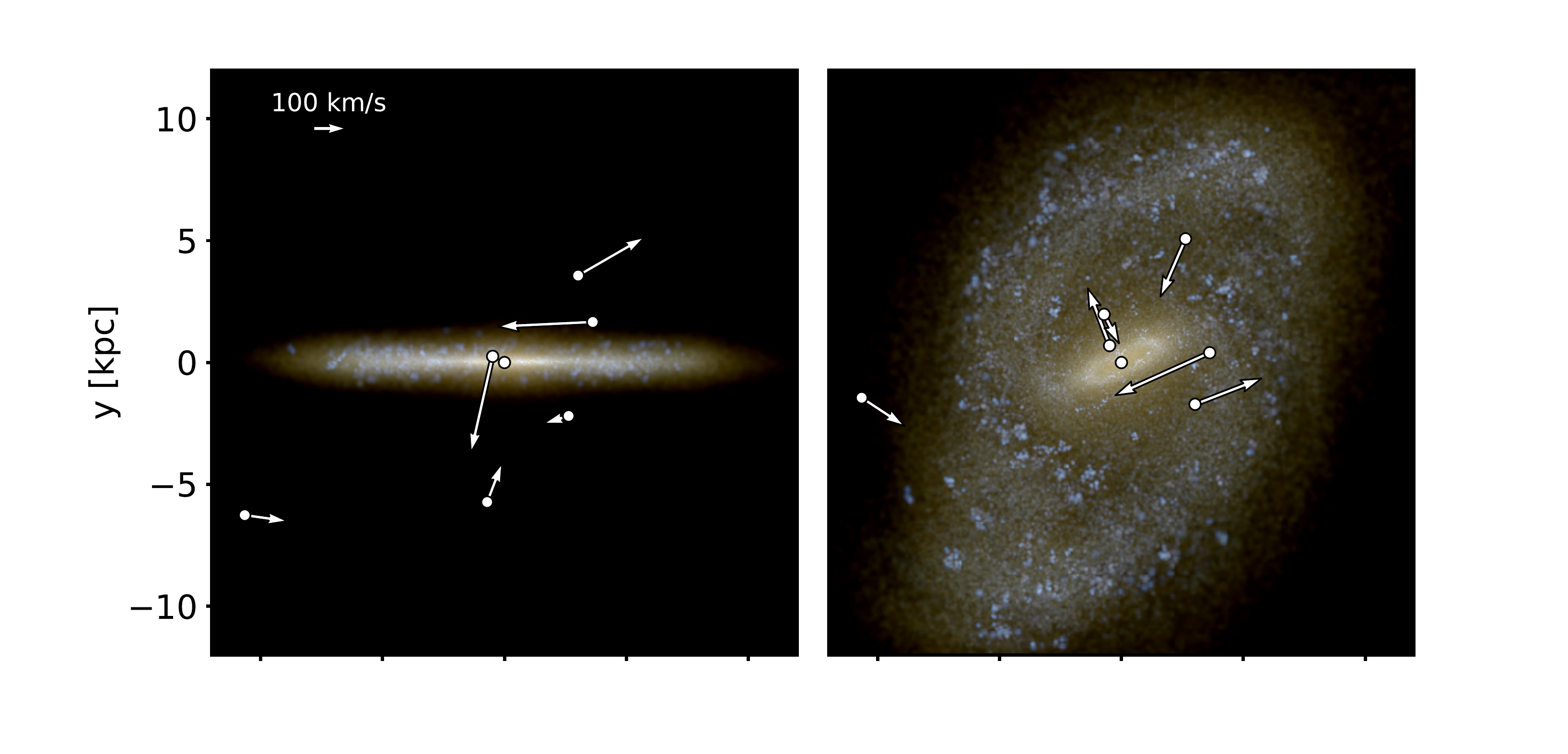}
\includegraphics[trim=15mm 17mm 20mm 15mm, clip, width=150mm]{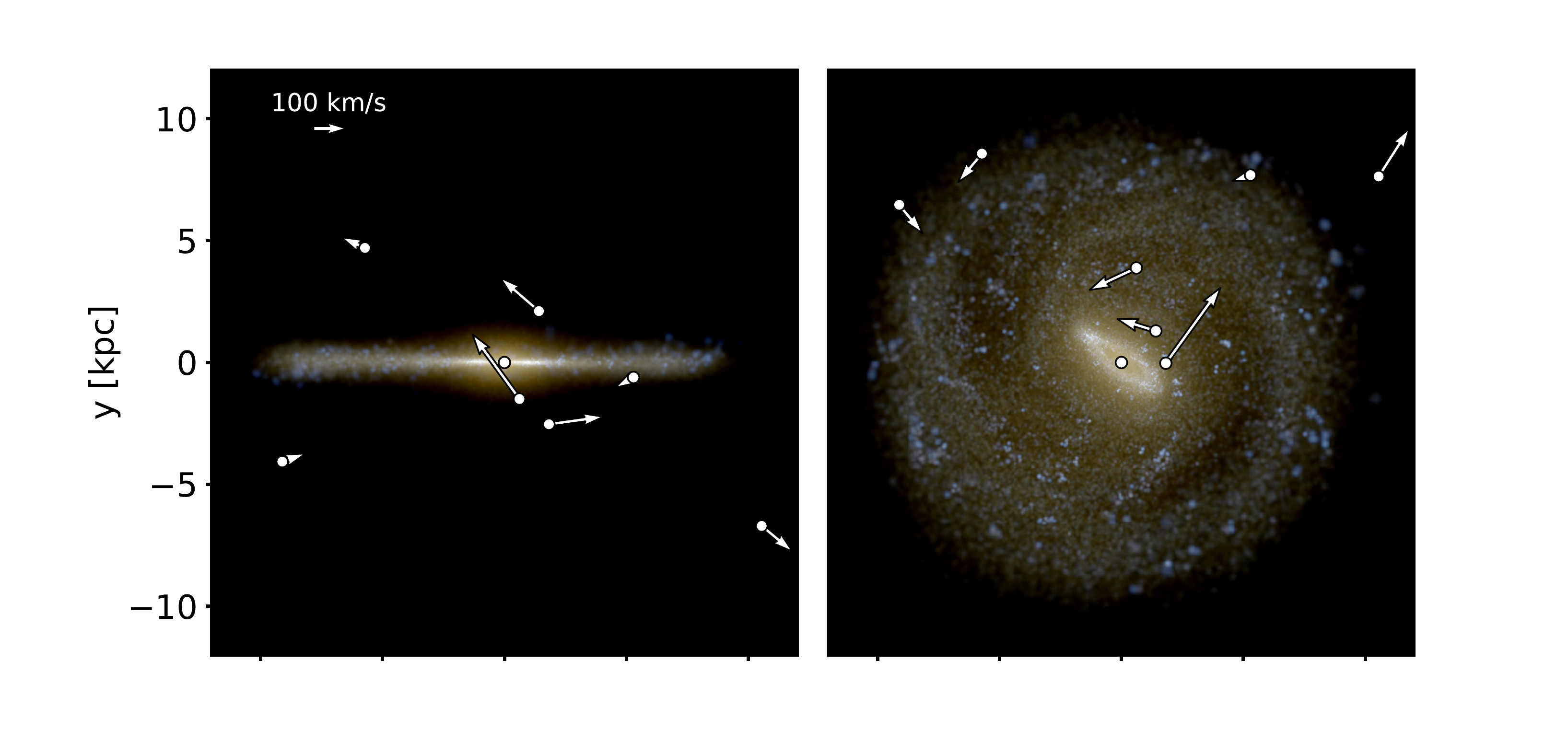}
\includegraphics[trim=15mm 0mm  20mm 15mm, clip, width=150mm]{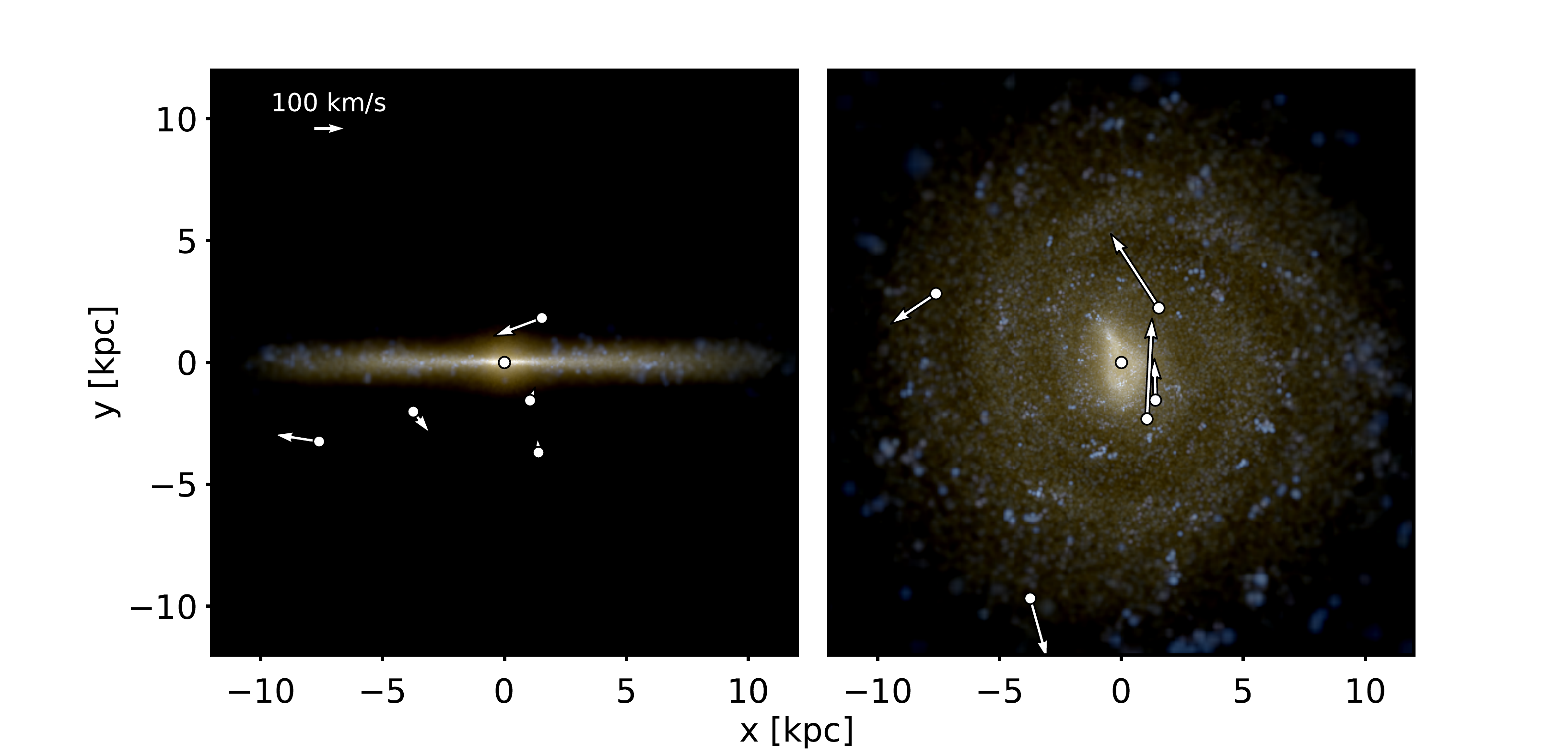}
\caption{{\sc Spatial Distribution and Velocities of Wandering SMBHs}. Stellar images of three galaxies from side-on (left) and face-on (right) orientations relative to the galactic disk. Galaxies were chosen from our parent sample of 26 galaxies in MW-mass halos to be most similar to the MW in terms of having a disk-dominated morphology and lacking any major (1:4 or larger) mergers since $z = 0.75$. Each pixel is colored based on emission in the U (blue), V (green), and J (red) bands, assuming a Kroupa IMF. Circles indicate position of the SMBHs and the arrows the direction and magnitude of their velocity relative to their galactic center. The wandering SMBHs have orbits with random inclination and eccentricity and are generally not in the galactic disk.}
\label{galaxy_images} 
\end{figure*}

We find a total of 316 SMBHs, including central SMBHs and excluding any within satellite halos, residing within the virial radius of 26 MW-mass halos in {\sc Romulus25}, an average of  $12.2 \pm 8.4$ SMBHs per halo and an average of $5.1 \pm 3.3$ SMBHs within 10 kpc of halo center, again including central SMBHs (133 total; errors are standard deviation). All but one of the 26 halos has a single central SMBH within the inner 0.7 kpc of the galaxy, with the one halo's most central SMBH residing approximately 1 kpc from halo center. Of the 108 offset SMBHs that exist within 10 kpc from halo center, 90\% entered into the inner halo ($D < 10$ kpc) more than 2 Gyr previously. Many have existed within their current host since the first few Gyr of the simulation. These are not SMBHs with orbits actively decaying toward the center, but SMBHs on long lived, kpc-scale orbits within their host galaxy.

In Figure~\ref{number_bhs} we plot the cumulative fraction of MW-mass halos hosting different numbers of SMBHs within their central 10 kpc, including central SMBHs. We select a subset of MW-mass halos that have had no mergers of total mass ratio greater than $0.25$ since $z = 0.75$ (16 halos) and another subset with central galaxies visually inspected to have a disk morphology at $z=0$ (12 halos). Wandering SMBHs are commonplace in MW-mass galaxies, with only one of 26 halos hosting just a single SMBH in its inner 10 kpc. The morphology or recent merger history of the galaxy does not affect the number of wandering SMBHs, consistent with the fact that many wanderers entered the galaxy at early times and are disconnected from its recent evolution.

\begin{figure}
\centering
\includegraphics[trim=10mm 7mm 10mm 30mm, clip, width=90mm]{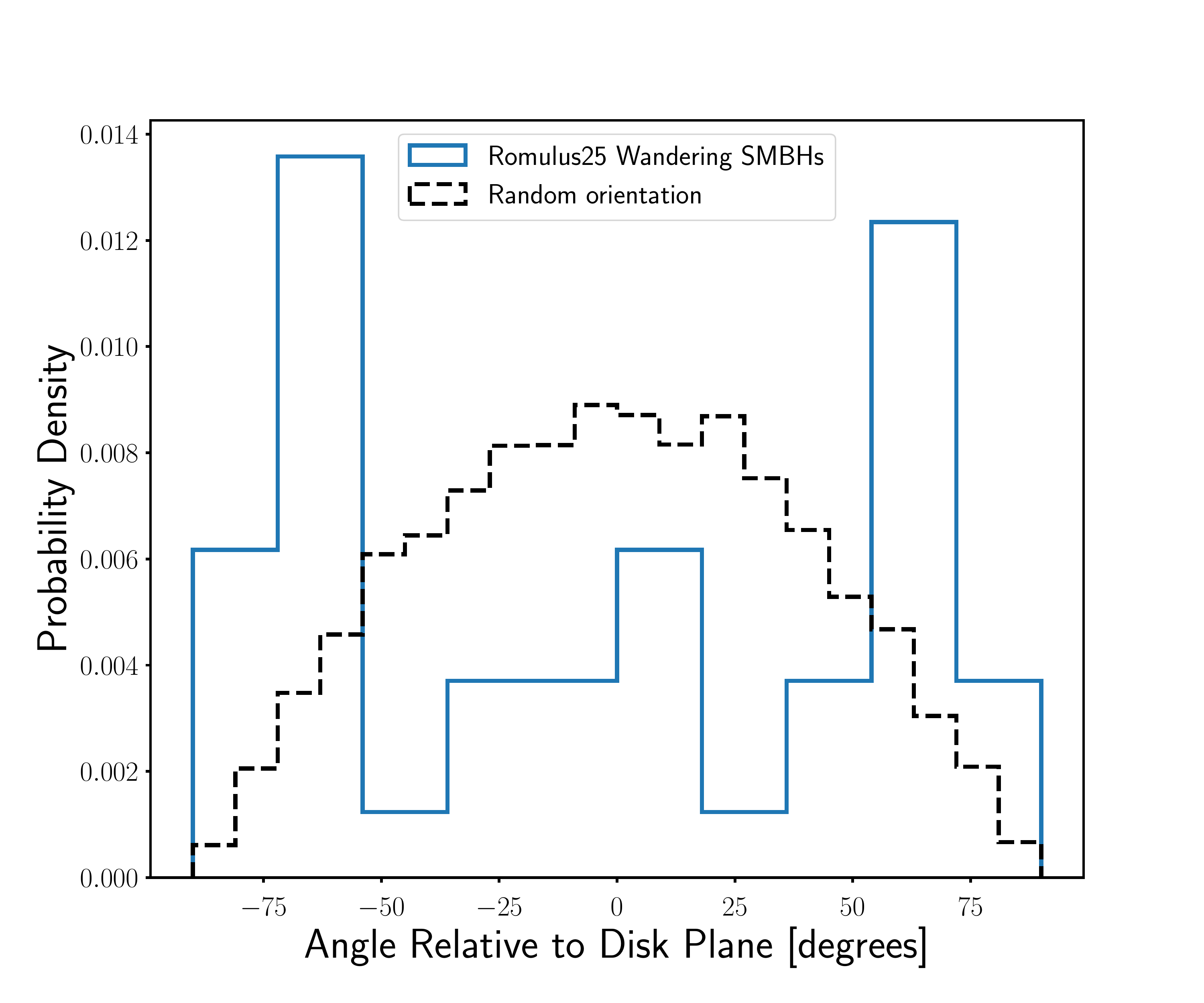}
\caption{{\sc SMBH Locations Within Disk Galaxies}. The distribution of polar angles relative to halo center and the the plane of the galactic disk for wandering SMBHs ($0.7 < D < 10$ kpc) in MW-mass disk galaxies from {\sc Romulus25} (blue line). The dashed black line represents the expectation from randomly sampling a unit sphere. In {\sc Romulus25} wandering SMBHs show a clear preference to exist out of the disk plane.}
\label{bh_angles} 
\end{figure}

Figure~\ref{galaxy_images} shows synthetic images of stars in three example galaxies, all of which are disk dominated and have had no recent major mergers, similar to our current model of the MW. SMBH orbits are generally not within the galactic disk and have a wide range of  inclinations and eccentricities. Out of the non-central ($D > 0.7$ kpc) SMBHs within 10 kpc of halo center in disk dominated galaxies, only $20\pm7\%$ exist within 30 degrees of the plane of the disk (see Figure~\ref{bh_angles}). If it were random, with the polar angle relative to the disk plane, $\phi$, taken from a flat distribution in $sin(\phi)$, 50\% of the SMBHs would be in this region, making this result significant to more than $4\sigma$. A significant fraction of these SMBHs have large velocities perpendicular to the disk, indicating that they are only passing through the region. This is consistent with previous work.  Mergers aligned with the galactic disk are more likely to deposit SMBHs closer to the galactic center \citep{callegariBH09,callegari11}. Further, the disk is denser with stars and gas, providing more efficient dynamical friction. 

The masses of wandering SMBHs are often very close to their initial mass, with over half of those within 10 kpc from halo center having grown by less than 70\% of their initial mass. The gas supply declines quickly away from the galactic center, making accretion more difficult to sustain. Still, some have grown by several times their initial mass, with 8\% having grown by over a factor of 10. Most of these SMBHs gain their mass through mergers and accretion early in the simulation before they are accreted onto their current host.  Others were once central SMBHs that got perturbed outward and replaced. In rare cases, wandering SMBHs are able to grow significantly while on orbits that take them periodically close to the galactic center. 

\begin{figure}
\centering
\includegraphics[trim=5mm 3mm 10mm 22mm, clip, width=90mm]{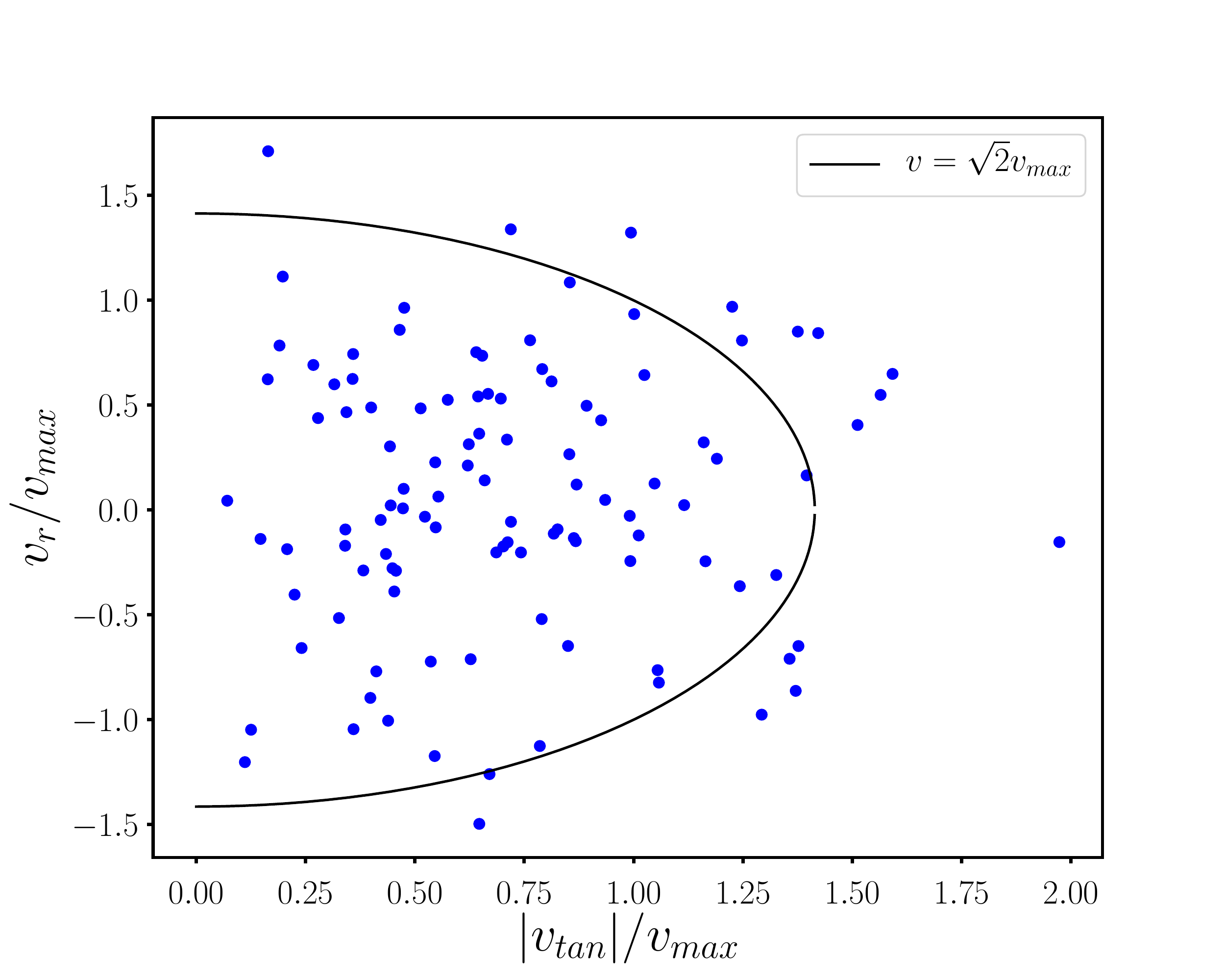}
\caption{{\sc Distribution of Wandering SMBH Velocities}. The radial velocity direction and magnitude versus total tangential velocity magnitude for wandering SMBHs ($D>0.7$ kpc) in the inner 10 kpc of MW-mass halos. Both velocities are given relative to the maximum circular velocity of their host halo,$v_{max}$. The solid black lines represent $\sqrt{2}v_{max}$. The orbits of wandering SMBHs have random eccentricities and the majority are bound to the central galaxy.}
\label{bh_velocities} 
\end{figure}

Figure~\ref{bh_velocities} shows the radial and tangential velocities of the SMBHs within 10 kpc of halo center in units of their host halos's maximum circular velocity, which varies between $\sim150-300$ km/s for our sample. The velocities are taken relative to the center of mass velocity of the halo's central kpc. The black line represents a total velocity of $\sqrt{2}v_{max}$, which is roughly equal to the escape velocity of the galaxy. Many of the SMBHs lie within this line. In all but two of the galaxies (which both show signs of disruption) the radius of maximum circular velocity is much less than 10 kpc (generally 1-3 kpc). This implies that most SMBHs are bound to the central galaxy while the 16\% that lie outside of this region are bound to the halo on larger scales and have more eccentric orbits.  

Some of these wandering SMBHs may be surrounded by dense nuclear star clusters with typical masses of 10$^{6-7}$ M${\odot}$ and effective radii of $\sim10$ pc \citep{wehner06,ferrarese06b,scottGraham13,scottGraham13b} . {\sc Romulus25} cannot resolve such detailed structures, but were they to exist they would effectively increase the dynamical mass of the SMBHs and potentially cause them to sink more efficiently. To test this effect, we follow the method used in \citet{Barausse12} and approximate the sinking timescale for each wandering SMBH using the formula derived in \citet{BinneyTremaine}.

\begin{equation}
\mathrm{t}_{\mathrm{df}} \sim \left( \frac{19\, \mathrm{Gyr}}{\ln\Lambda} \right ) \left( \frac{r_i}{5\, \mathrm{kpc}} \right ) \left ( \frac{\sigma}{200\, \mathrm{km/s}} \right ) \left( \frac{10^8\, \mathrm{M}_{\odot}}{m_{\mathrm{bh}}} \right ) \\
\end{equation}

\noindent We assume $\Lambda \sim b_{max}/b_{min}$ with the maximum impact parameter, $b_{max}$, equal to the initial radius of the SMBH orbit, $r_i$. We take the velocity dispersion of each halo to follow the empirical relations from \citet{oser12}.

\begin{equation}
\sigma \sim 190 \left ( \frac{\mathrm{M}_{\star}}{10^{11} \mathrm{M}_{\odot}} \right )^{0.2} (1+z)^{0.44}
\end{equation}

Using Eq. (1) with $r_i$ equal to the final radial distance of each SMBH from galaxy center, we estimate the dynamical friction timescale at $z = 0$ for each wandering SMBH in the inner 10 kpc of their host halo when a factor of 10 is added to their mass (a high estimate of the ratio of nuclear star cluster to SMBH mass). We apply a minimum impact parameter, $b_{min}$, of $10$ pc to represent the characteristic size of nuclear star clusters. While this significantly decreases the timescale estimated by Eq. (1), 65\% of these wandering SMBHs still have a sinking timescale longer than the total time they've spent at $D < 10$ kpc (50\% for $D < 5$ kpc). While this is a simplistic estimate, it indicates that our results are robust to the existence of unresolved, massive stellar components around SMBHs.

The orbits of SMBH pairs are not followed at separations closer than 0.7 kpc (see \S2), but we test whether it is likely that unresolved wandering SMBHs exist at smaller scales using Eqs. (1) and (2). We take the initial orbital radius, $r_i$, to be the smallest resolved separation, 0.7 kpc. Because Eq. (2) is fit only to galaxies at $z<2$ we evaluate the equation with $z = \mathrm{min}(z_{pair},2)$, where $z_{pair}$ is the redshift of close pair formation. We take $b_{max}$ to be 0.7 kpc and $b_{min}$ to be  the 90 degree deflection radius of the SMBH. The stellar mass is taken directly from the pair's host galaxy at the appropriate redshift. Whether calculated with host properties at $z_{pair}$ or at $z=0$, the results of Eqs. (1) and (2) predict that each central close pair should form a bound binary well before the Hubble time. Because {\sc Romulus25} does not include lower mass black holes that would have much longer binary formation timescales, these results are not in tension with recent claims for black holes with mass $\sim10^4$ M$_{\odot}$ on sub-kpc orbits in the MW \citep{oka17,tsuboi17}.





\begin{figure}
\centering
\includegraphics[trim=17mm 5mm 10mm 20mm, clip, width=90mm]{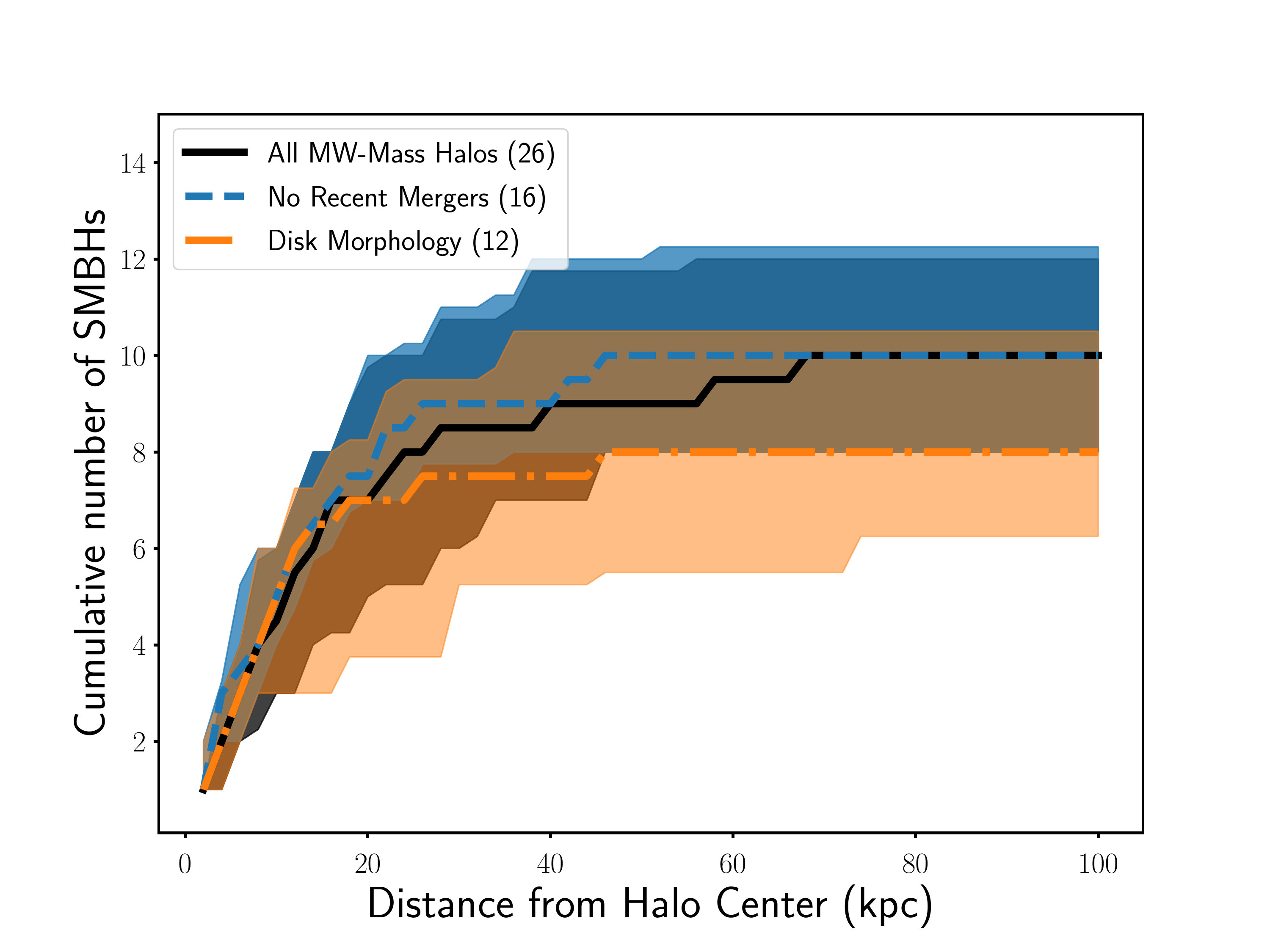}
\caption{{\sc Distance Distribution for Wandering SMBHs}. The median cumulative number of wandering SMBHs as a function of distance from halo center for MW-mass halos. Shaded regions span 75th and 25th percentiles. The colors and styles are the same as Figure~\ref{number_bhs}.}
\label{distances} 
\end{figure}

Figure~\ref{distances} plots the median cumulative distribution of SMBH distances in MW-mass halos. Many wandering SMBHs exist at $R>10$ kpc. These SMBHs have grown even less than those closer to halo center, with a median mass of only 1.07 times their initial seed mass and 70\% having grown by less than a factor of 2. There is again not any clear dependence on galaxy morphology or recent merger history.


\section{Summary}
We present a prediction from a large-scale cosmological simulation for the population of wandering SMBHs in MW-mass halos using 26 halos extracted from the {\sc Romulus25} simulation. {\sc Romulus25} is uniquely able to track the evolution of SMBH orbits during and following galaxy mergers while self-consistently accounting for the changing kinematics and structure of their host galaxy \citep{tremmel15,tremmel17,tremmel18}.


We predict that MW-mass halos often host several wandering SMBHs on kpc-scale orbits regardless of merger history or morphology. Wandering SMBHs are unlikely to be found in galactic disks, a result that is consistent with previous works \citep{callegariBH09,callegari11}. The majority of wandering SMBHs have grown little since their seeding, though some have grown substantially, with growth mostly occurring at early times prior to their arrival in their final host. Previous works have shown how SMBHs are commonly deposited on wide, long lived orbits within galaxies as a result of the disruption of their host during a galaxy merger \citep{Yu02,callegariBH09,callegari11,dosopoulou17,tremmel18}. Without any stellar core to assist in their orbital decay, the SMBHs will remain on kpc-scale orbits for long periods of time. The merger history of MW-mass halos, including our own, have included many minor mergers more likely to lead to tidal disruption and the deposition of `naked' SMBHs within the galaxy \citep{tremmel18}.

Such a population of wandering SMBHs in MW-mass halos has been predicted both by semi-analytic models \citep[e.g.][]{volonteriOffCenBH05,dvorkin17} as well as cosmological simulations \citep{bellovaryBH10,volonteri16}. An improvement over previous cosmological simulations, {\sc Romulus25} is able to accurately track the orbital evolution of SMBHs using a well tested technique that accounts for their orbital decay due to unresolved dynamical friction \citep{tremmel15}. SMBHs are seeded in very specific environments at the centers of proto-galaxies at high redshift. The results presented here self-consistently incorporate galaxy merger histories as well as SMBH occupation and dynamical evolution with fewer \textit{a priori} assumptions and in more detail compared to previous cosmological simulations and semi-analytical models. {\sc Romulus25} does not include the effect of three body interactions, nor does it include prescriptions for gravitational recoil events resulting from SMBH mergers. Both may contribute further to the population of wandering SMBHs in galaxies \citep[e.g.][]{volonteriOffCenBH05,blecha16}. 

Observing wandering SMBHs in massive galaxies can provide unique constraints on SMBH formation and early growth. This may be possible in the near future if these SMBHs retain a bound stellar population around them, if they interact with nearby stars and result in tidal disruption events, or if they accrete gas and appear as an ultra-luminous X-ray source or off-center active galactic nucleus. We will explore the observable nature of wandering SMBHs in future work.
\\

 \section*{Acknowledgments}
FG, TQ and MT were partially supported by NSF award AST-1514868.  MT gratefully acknowledges
support from the YCAA Prize Postdoctoral Fellowship. AP was supported by the Royal Society. This research is part of the Blue Waters sustained-petascale computing project supported by the National Science Foundation (awards OCI-0725070 and ACI-1238993) and the state of Illinois. Blue Waters is a joint effort of the University of Illinois at Urbana-Champaign and its National Center for Supercomputing Applications. This work is also part of a PRAC allocation support by the National Science Foundation (award number OCI-1144357). MV acknowledges funding from the European Research Council under the European Community's Seventh Framework Programme (FP7/2007-2013 Grant Agreement no. 614199, project `BLACK'). The analysis was done using software packages Pynbody \citep{pynbody} and TANGOS \citep{tangos}. The authors thank Jillian Bellovary, Priyamvada Natarajan, and Angelo Ricarte for a thorough reading of the manuscript and useful discussion.
\bibliographystyle{apj}

\end{document}